# Beam based calibration of X-ray pinhole camera in SSRF[*]

LENG Yong-Bin（冷用斌）[1,2][†]   HUANG Guo-Qing（黄国庆）[1,2]   ZHANG Man-Zhou（张满洲）[1,2]
CHEN Zhi-Chu（陈之初）[1,2]   CHEN Jie（陈杰）[1,2]   YE Kai-Rong（叶恺容）[1,2]

[1] *Shanghai Synchrotron Radiation Facility, Shanghai 201800, China*
[2] *Shanghai Institute of Applied Physics, Shanghai 201800, China*

The Shanghai Synchrotron Radiation Facility (SSRF) contains a 3.5-GeV storage ring serving as a national X-ray synchrotron radiation user facility characterized by a low emittance and a low coupling. The stability and quality of the electron beams are monitored continuously by an array of diagnostics. In particular, an X-ray pinhole camera is employed in the diagnostics beamline of the ring to characterize the position, size, and emittance of the beam. The performance of the measurement of the transverse electron beam size is given by the width of the point spread function (PSF) of the X-ray pinhole camera. Typically the point spread function of the X-ray pinhole camera is calculated via analytical or numerical method. In this paper we will introduce a new beam based calibration method to derive the width of the PSF online.

**Keywords:** Beam diagnostics; X-ray pinhole camera; point spread function; beam size measurement.

**PACS:** 29.20.db; 29.27.Fh; 29.85.Fj.

## 1 Introduction

The transverse beam emittance is a crucial accelerator parameter because it is directly related to the brilliance of a synchrotron light source. Due to its non–destructive nature synchrotron radiation from a bending magnet is a versatile tool for beam profile measurements and is used in nearly every accelerator[1]. The spatial resolution better than a few microns is now demanded for transverse beam profiling on the most performing light sources. For this purpose X-ray pinhole cameras are widely used due to simple setup and high practical reliability[2–5]. Typically a pinhole based emittance monitor has a limited resolution of $\geqslant 10~\mu$m. Recent calculations taking into account the spectral distribution of the source and applying numerical methods to precisely evaluate the diffraction, have shown that with the correct choice of pinhole size and magnification a better resolution can be achieved. Precise calculation or calibration of the point spread function of the whole system is required in this case[6 and 7].

As a third generation national X-ray light source, the Shanghai Synchrotron Radiation Facility (SSRF) storage ring must be held to a high performance level[8]. An X-ray pinhole camera is implemented to monitor beam position and measure beam size and emittance precisely. The point spread function of the X-ray pinhole camera is calculated with analytical method in the design stage. In order to determine the practical value of point spread function a beam experiment has been carried out. The details and results of this beam based calibration will be discussed in this paper.

[*] Supported by NSFC (11075198)
[†] lengyongbin@sinap.ac.cn

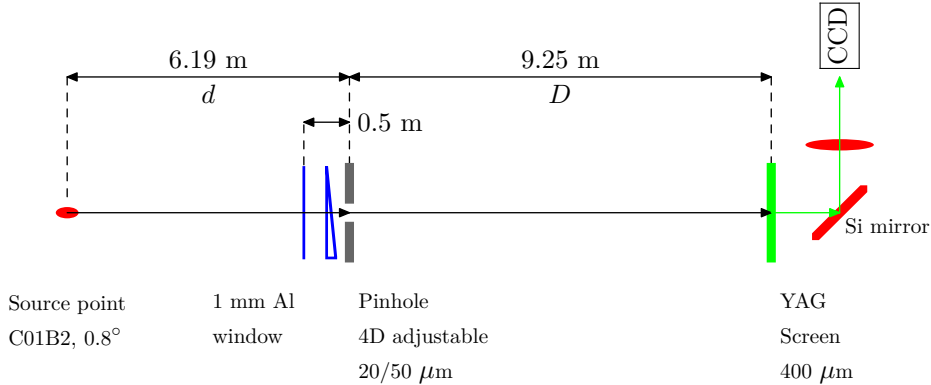

**Figure 1**  System layout of SSRF X-ray pinhole camera.

## 2  System setup

### 2.1  SSRF Xray Pinhole layout

The basic layout and components of the SSRF pinhole camera are shown in **Figure 1**.

The X-ray beam from the bending magnet goes an aluminum window which transmits only the high energy photons from vacuum to air. A pinhole array combined by two sets of tungsten slits, in horizontal and vertical directions, is placed behind the window, as close as possible from the source, 6.19 m in our case. The X-ray image of the pinhole camera was converted into a visible light image at peak wavelength of 530 nm with an YAG screen (400 $\mu$m thick) . The screen is placed at 9.25 m away from the pinhole, so that the image is magnified by a factor of 1.5. To acquire and measure the size of the source we image the screen with a macro-lens (Componon 2.8/50 from Schneider-Kreuznach) focusing on a compact IEEE 1394 CCD camera (AVT Guppy F-080B, pixel size 4.65 $\mu$m) A LabVIEW and Shared Memory IOCcore technique based application has been developed to control the camera and communicate with control system through EPICS CA protocol. The $x$, $y$ position of the CCD, the $x$, $y$ position and angles of pinholes array can be remotely adjusted[9].

### 2.2  PSF calculation

The performance of the measurement of the transverse electron beam size is given by the width of the point spread function (PSF) of the X-ray pinhole camera. The image formed on the camera is the convolution of several independent contributions including beam size, pinhole size, YAG screen and CCD. Let us call $\Sigma$ the r.m.s. Gaussian size of the acquired image and assume the source and the PSF's to be Gaussian. Then $\Sigma$ can be expressed as follow:

$$\Sigma = \left[(S \cdot C_{\mathrm{mag}})^2 + S_{\mathrm{aper}}^2 + S_{\mathrm{diff}}^2 + S_{\mathrm{scr}}^2 + S_{\mathrm{CCD}}^2\right]^{1/2} = \left[\left(S\frac{D}{d}\right)^2 + S_{\mathrm{sys}}^2\right]^{1/2}, \qquad (1)$$

where $S$ is the r.m.s. size of the image of the photon emitted by electron beam at the source point, $S_{\mathrm{aper}}$ is the geometrical contribution introduced by the finite size of the pinhole, $S_{\mathrm{diff}}$ is the diffraction contribution by the small pinhole, $S_{\mathrm{scr}}$ is the screen spatial resolution, $S_{\mathrm{CCD}}$ is the spread induced by the camera, which includes pixel size, lens aberration and depth of focus through the finite thickness of the screen and aperture of the lens and $S_{\mathrm{sys}}$ is effective PSF of the whole pinhole system. $C_{\mathrm{mag}}$ is the magnification factor of the pinhole camera. $d$ is the distance from source point to pinhole. $D$ is the distance from pinhole to screen.

Let $A$ be the aperture of a rectangular shaped pinhole. The contribution of the diffraction for a monochromatic photon beam of wavelength $\lambda$ is given analytically by

$$S_{\text{diff}} = \frac{\sqrt{12}}{4\pi}\frac{\lambda D}{A}. \quad (2)$$

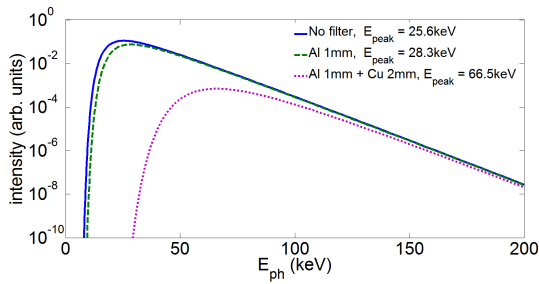

**Figure 2** Spectrum of extracted X-ray photon beam.

A simple geometrical computation shows that $S_{\text{aper}}$ can be expressed as[10]:

$$S_{\text{aper}} = \frac{A}{\sqrt{12}}\frac{D+d}{d}. \quad (3)$$

For the calculation of the PSF of the pinhole, the spectrum of the source needs to be taken into account. In our case the spectrum is the synchrotron radiation from a bending magnet, filtered in energy and intensity by a 1 mm thick Al window, 2 mm thick Cu filter and 9.7 m of air. **Figure 2** shows the photon beam spectra at source point, after 1 mm Al window and after 2 mm Cu filter calculated by XOP[‡] software package.

Applying the practical value of $D$ 6.19 m, $d$ 9.25 m and $A$ 50.0 $\mu$m, we can get $S_{\text{aper}}$ to be 36.0 $\mu$m. Integrating the expression **(2)** over the spectrum, we can get $S_{\text{diff}}$ to be 0.9 $\mu$m. Then the r.m.s. combination of $S_{\text{aper}}$ and $S_{\text{diff}}$ is determined to be 36.0 $\mu$m for 50.0 $\mu$m pinhole at SSRF.

Sscreen and SCCD are hard to be calculated analytically or numerically and have to be derived by experiment. In our case the width of PSF of $S_{\text{scr}}$ and $S_{\text{CCD}}$, which is about 10.0 $\mu$m for 400 $\mu$m YAG screen, is adopted from experimental data of Diamond Light Source in the design stage[6].

Totally the effective PSF of the whole pinhole system Ssystem is calculated to be 37.4 $\mu$m with the above configuration in the design stage.

## 3 Beam based calibration

In order to verify the usability and reliability of pinhole camera and determine the width of the system point spread function, a new beam based calibration method had been developed in the SSRF storage ring. By varying the beam size $S$ at the source point and measuring image size $\Sigma$, the practical $S_{\text{sys}}$ can be derived from equation **(1)** using least-square fitting method.

After the linear optics measurements and optimization procedure known as LOCO, the maximum beta function beating of SSRF storage ring had been minimized to smaller than 1 %. In this case the difference of beam parameters between the model and practical machine was smaller than 1 % too. We could use the beam size value of model $S_{\text{model}}$ to replace practical beam size value $S$.

As we know the beam size could be described as follow:

$$S_i^2 = \beta_i \epsilon_i + (\eta_i \sigma_\epsilon)^2, \quad (4)$$

where $S_i$ is the beam size in the horizontal or vertical plane respectively ($i = x, y$), $\beta_i$ and $\eta_i$ are the betatron and dispersion functions at the source point and in the corresponding plane; and $\epsilon_i$ and $\sigma_\epsilon$ are the emittance and the relative energy spread of the electron beam.

The experiment of PSF calibration was carried out in horizontal plane with 500 bunches 170 mA electron beam. The beam size of source point was changed by modifying the power supply current $I_{Q5}$ of the 5th set of quadrupoles. The measured image sizes $\Sigma_x$ at CCD plane were recorded for each $I_{Q5}$ setting. Applying the machine parameters ($\epsilon_x$, $\beta_x$, $\eta_x$ and $\sigma_\epsilon$), which were derived from optimized model corresponding to different $I_{Q5}$ settings, to equation

---

[‡] http://www.esrf.eu/UsersAndScience/Experiments/TBS/SciSoft/xop2.3/Main

| $I_{Q5}$ (A) | $\epsilon_x$ (nm·rad) | $\beta_x$ (m) | $\eta_x$ (m) | $S_x$ ($\mu$m) | $\Sigma_x$ ($\mu$m) |
| --- | --- | --- | --- | --- | --- |
| 0.94 | 6.54 | 1.04 | 0.084 | 130.2 | 184.0 |
| 0.95 | 5.61 | 0.99 | 0.076 | 117.8 | 167.2 |
| 0.96 | 4.84 | 0.94 | 0.069 | 106.8 | 153.7 |
| 0.97 | 4.46 | 0.90 | 0.063 | 98.6 | 142.5 |
| 0.98 | 4.15 | 0.86 | 0.058 | 91.7 | 133.3 |
| 0.99 | 3.97 | 0.83 | 0.053 | 85.7 | 125.9 |
| 1 | 3.91 | 0.79 | 0.049 | 80.9 | 120.0 |
| 1.01 | 3.93 | 0.76 | 0.045 | 76.8 | 116.9 |
| 1.02 | 4.03 | 0.73 | 0.041 | 73.2 | 112.7 |
| 1.03 | 4.21 | 0.70 | 0.039 | 71.7 | 109.5 |
| 1.04 | 4.44 | 0.68 | 0.035 | 69.2 | 106.9 |
| 1.05 | 4.73 | 0.65 | 0.032 | 67.4 | 105.1 |
| 1.06 | 5.07 | 0.62 | 0.029 | 66.0 | 103.7 |
| 1.07 | 5.46 | 0.58 | 0.025 | 63.8 | 102.7 |

**Table 1** The expected beam sizes and the corresponding Guassian image sizes obtained from various machine parameters.

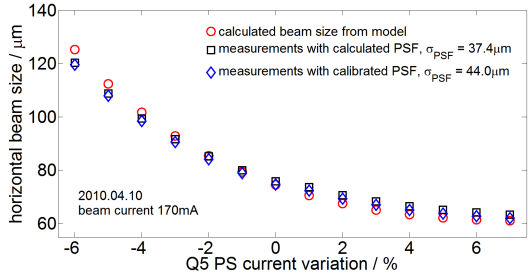

**Figure 3** Comparison of calculated beam size and measured beam size.

**(4)**, the expected beam sizes $S_x$ could be obtained as **Table 1**. In this calculation the energy spread $\sigma_\epsilon$ of 0.0011 was used based on the previous experiment.

**Figure 3** shows the beam size comparison of calculation and measurements. It is obvious that the experimental data has good agreement with modeling data, which verifies the good usability and reliability of SSRF X-ray pinhole camera.

Fitting $\Sigma_x$ and $S_x$ data with equation **(1)** using least-square method the effective width of pinhole camera PSF had been determined to be 44.0 $\mu$m larger than expected value 37.4 $\mu$m.

There were two probable causes of this disagreement. The first one was that Gaussian PSF assumption was not perfectly suitable for large size pinhole (50 $\mu$m in this case). The second one was that the 10 $\mu$m width of PSF of $S_{\text{scr}}$ and $S_{\text{CCD}}$, adopted from Diamond Light Source, was not perfectly suitable for SSRF pinhole setup.

## 4 Applications

The beam based calibration experiment showed the micron level sensitivity of SSRF pinhole camera. After applying the new calibrated PSF the system performance has been improved to be qualified as a precise beam status monitor. The differences between measured beam sizes and modeling beam sizes decreased from 5 % to smaller than 4 % after beam based calibration.

Presently the pinhole camera is routinely used as both a beam imaging device to monitor transverse distribution and as an emittance diagnostic. During user runs, the SSRF storage ring horizontal beam size at the bending magnet is normally stable within a 2 $\mu$m range (2.9 %) shown in **Figure 4(a)**. The beam size disturbance of horizontal beam size induced by injection has been observed as small as 0.5 $\mu$m

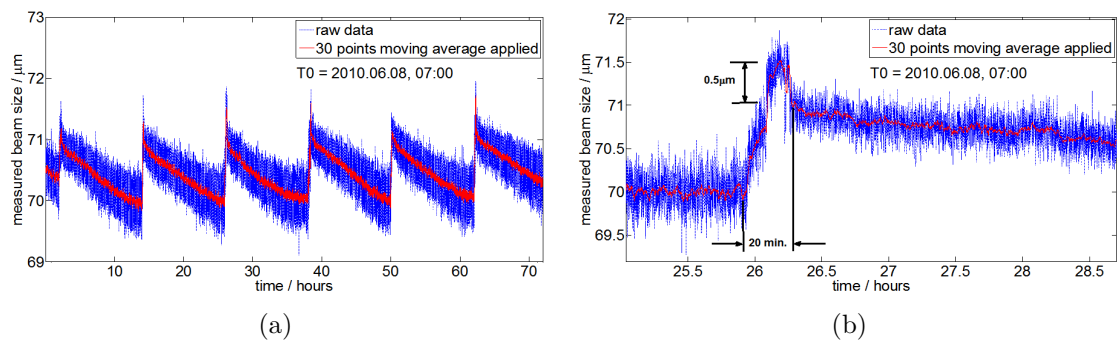

**Figure 4** (a) 70 hours history data of horizontal beam size. (b) Beam size disturbance due to injection shown in **Figure 4(b)**.

## 5 Conclusion

An X-ray pinhole camera diagnostic has been built at the SSRF. To determine the width of system PSF for the pinhole camera a beam based calibration method has been introduced. This calibration allows us to derive a more accurate resolution for the system. The variation of horizontal beam size between theoretical value and measured value is as small as 4 % after calibration.

# 基于束流的上海光源 X 射线针孔相机标定[**]

冷用斌[1,2][††]   黄国庆[1,2]   张满洲[1,2]   陈之初[1,2]   陈杰[1,2]   叶恺容[1,2]

[1] 上海光源,
[2] 上海应用物理研究所

上海光源是一台能量为 3.5 GeV 的低发射度、低耦合度的国家同步辐射光源。同步辐射光源中电子束流的品质必须采用一系列的束流诊断设备进行在线监测。在专用诊断线站上建立 X 射线针孔相机对束流位置、束斑尺寸及发射度进行测量，是近年来常用的解决方案。对针孔相机而言横向束斑尺寸的测量精度主要由系统的点扩散函数（PSF）确定，其数值通常采用解析推导或是数值计算的方法得到。本文介绍了一种新的基于束流的点扩散函数在线标定方法。

**Keywords:** 束流诊断; X 射线针孔相机; 点扩散函数; 束斑尺寸测量.

**PACS:** 29.20.db; 29.27.Fh; 29.85.Fj.